\documentclass[preprint]{epl}
\usepackage{epsfig}
\usepackage{graphicx}
\usepackage{amsmath}
\usepackage{color}
\title{Kinetics of depletion interactions}
\author{ G. A. Vliegenthart\inst{1,2}\thanks{Corresponding author: g.vliegenthart@fz-juelich.de.} \and P. van der Schoot\inst{3} }
\shortauthor{G. A. Vliegenthart \etal}
\institute{ 
\inst{1} Institut for Festk\"{o}rperforschung - Forschungszentrum J\"{u}lich, D-52425 J\"{u}lich, Germany\\
\inst{2} School of Chemistry - University of Bristol, Cantocks Close, Bristol, BS8 1TS, United Kingdom\\
\inst{3}Department of Applied Physics - Eindhoven University of Technology,
P.O. Box 513, 5600 MB Eindhoven, The Netherlands}
\pacs{82.70.Dd}{Colloids}
\pacs{05.20.Dd}{Kinetic theory}
\pacs{05.10.Gg}{Stochastic analysis methods}

\begin{document}
\maketitle 

\begin{abstract}
Depletion interactions between colloidal particles dispersed in a fluid medium are effective interactions induced by the presence of other types of colloid.  They are not instantaneous but build up in time. We show by means Brownian dynamics simulations that the fluctuations in the depletion force between two  guest particles in a host dispersion of differently sized colloids do not decay exponentially with time, but show a power-law dependence. A simple scaling theory  accurately describes the dependence of the magnitude of these fluctuations on time, on the inter-particle distance and on the size ratio of guest and host particles. The consequences in particular for the dynamics of  colloidal mixtures are discussed.
\end{abstract}

\section{Introduction}
A useful tool in the equilibrium theory of colloidal mixtures is the `integrating out' of the degrees of freedom of one of the components \cite{DIJKSTRA99}. The resulting renormalisation of the interaction potential between the remaining particles in what has become an effective one-particle description, is often referred to as the depletion potential. 
Depletion interactions play a central role in the current understanding of the phase behaviour of dispersions containing colloids of different size and/or shape and have a remarkably elegant physical interpretation, although strictly accurate only in the dilute limit. This interpretation hinges on the view that an imbalanced osmotic pressure pushes two test particles towards each other when they approach within a correlation length set by the other species \cite{ASAKURA58}. 

Depletion potentials are often treated as true, instantaneous potentials even when considering time-dependent phenomena such as the kinetics of demixing and gelation \cite{BODNAR96,VERHAEGH96,MELROSE99,POON99}. This is problematical as depletion interactions need time to build up  through many collisions between a pair of test particles and the particles of the other species. This is (also) the reason why both in experiments \cite{RUDHARDT98,VERMA98} and in computer simulations \cite{LOEWEN00} the (mean) depletion potential is only obtained after extensive averaging. Here we show by means of scaling theory and Brownian dynamics simulations that the static depletion interaction potential is only approached algebraically in time.  Clearly, this must have an impact upon processes such as phase-separation kinetics and gelation. 

In this letter we first present the scaling theory of the decay of the fluctuations in the depletion force. Next, we outline the simulation technique used and present the results of our simulations. We end the paper with concluding remarks.

\section{Scaling theory}
Arguably, diffusive processes are at the root of the buildup of depletion forces, allowing us to estimate the kinetics thereof by fairly simple scaling arguments. For the sake of  simplicity we consider the situation of two parallel plates in a dispersion of ghost-like particles,  which do not interact amongst themselves but only with the plates through volume exclusion. Our arguments carry over straightforwardly to the more interestingly case of pairs of spheres, the results of which we give at the end of this section.

Considered are two parallel plates of linear dimension $\sigma_1$ separated by a distance $R$ immersed in a dispersion of (ghost) particles of diameter $\sigma_2$ at a number density $\rho_2$.  The presence of the ghost spheres induce a net interaction as they are excluded from the gap between the plates when $R$ is smaller than $\sigma_2$. According to the Asakura-Oosawa depletion theory,  \cite{ASAKURA54} the mean interaction force between the plates is given by $\left<F\right>=\Pi \sigma_1^2$ for all separations $R \leq \sigma_2$, where $\Pi = k_BT\rho_2$ is the osmotic pressure of the host dispersion and $k_BT$ the thermal energy. For $R>\sigma_2$, $\left<F\right>=0$.

The mean number of ghost colloids $N_2(t)$ contributing to the net force $\left<F\right>$ between the plates is equal to their mean density times the plate area times a diffusion length, and increases with time  proportionally to  $\rho_2 \sigma_1^2 \sqrt{D_2 t}$ with $D_2$ the diffusivity of species 2 and $t$ the time. Assuming the ideal host dispersion to be in local equilibrium, it follows from standard statistical mechanics that the mean-square fluctuation of this number must also be of the order ${N_2(t)}$. As a consequence, the mean-square fluctuation $\left<\Delta F^2\right>$ of the depletion force should scale as  $\rho_2 k_BT \sigma_1^2 / \sqrt{D_2 t}$. Using the Stokes-Einstein relation $D_2\sim k_BT/\mu \sigma_2$ for the diffusivity of colloid species 2,  with $\mu$ the viscosity of the solvent, we thus obtain
\begin{equation}
\left<\Delta F^2\right> \sim \rho_2 (k_BT)^{3/2} \sigma_1^2  \sigma_2^{1/2}   \mu^{1/2}  t^{-1/2}\label{eq:fluctplate}
\end{equation}
for $R<\sigma_2$, showing an algebraic decay with time as advertised in the introduction. This is due to our assumption of the diffusive nature of the coupling of the plates to the host dispersion. On account of the fluctuation-dissipation theorem, eq.~(\ref{eq:fluctplate})  implies that the buildup of the depletion forces  must also be algebraic in time. From eq.~(\ref{eq:fluctplate}) we furthermore conclude that for $t \gg \tau \equiv \sigma_2  \mu /\rho_2^2  (k_BT)^3 \sigma_1^4$ the depletion interaction between the plates has saturated to its mean-field value. The plates act as a microscopic pressure gauge, which equilibrates faster, i.e., has a smaller crossover time $\tau$, the larger the plates, the higher the temperature, the higher the density of the ghost particles, et cetera, as one would in fact expect. 
 
\begin{figure}
\begin{center}
\includegraphics[scale=0.7]{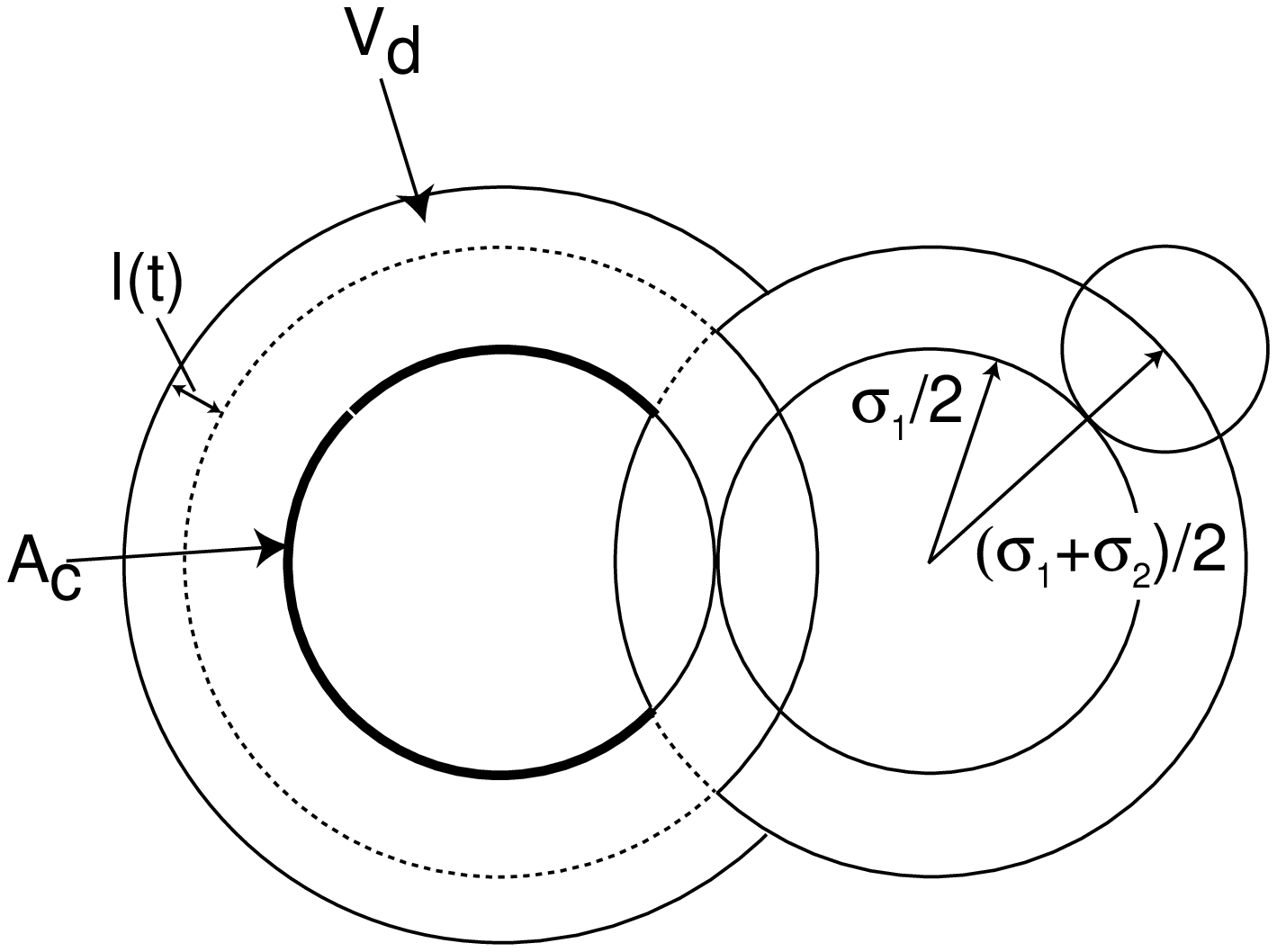}
\caption{Schematic illustration of the diffusion volume $V_d$ and the collision area $A_c$ for two spheres. The diffusion volume is  defined as a spherical shell of thickness $l(t)$ around a guest particle from which the host particles can diffuse towards the surface for collision. The collision area is the available area for collisions with a host particle taking into account momentum transfer only along the vector connecting the centers of mass of the two guest particles. }\label{fig:nuts}
\end{center}
\end{figure}

Our arguments carry over naturally to physically more interesting geometries such as pairs of large spheres immersed in dispersions of (ghost) spheres of different size, although that the available collision area and the diffusion volume are no longer trivial to evaluate, because they now depend on the size ratio of the spheres as well as on their separation (see fig.~\ref{fig:nuts}).  Here, we simply reproduce without proof for that particular case the final results for the mean and the mean-square fluctuation of the depletion force. The former reads $ \left<F\right>=\pi \sigma_1^2(1+q)^2[1-R^2/\sigma_1^2(1+q)^2]\rho_2 k_BT$ for $\sigma_1 < R \leq \sigma_1+\sigma_2$ and $ \left<F\right>=0$ for $R>\sigma_1+\sigma_2$, where $q=\sigma_2/\sigma_1$ is the size ratio.
The latter can be shown to obey
\begin{equation}
\left<\Delta F^2\right> \sim \frac{(1+q)[  \sigma_1^2(1+q)^2[1+R^2/\sigma_1^2(1+q)^2] k_BT ]^2\rho_2}{ l(t)[(1+q)+R/\sigma_1][3\sigma_1^2(1+q)^2/4+3\sigma_1 (1+q)l(t)/2 + l(t)^2]}\label{eq:fluctsphere}
\end{equation}  
for $\sigma_1 < R \leq \sigma_1+\sigma_2$, where $l(t)=c\sqrt{D_2 t}$ is a diffusion length. Here, $c$ is an  unknown constant of proportionality that we fix by fitting to the results of the Brownian dynamics simulations.  Equation \ref{eq:fluctsphere} holds also for $R\geq\sigma_1+\sigma_2$ provided we set $R=\sigma_1+\sigma_2$; the magnitude of the fluctuations in the depletion force level off to that  of the net zero force on a single sphere, if the separation of the spheres is larger than the diameter of the ghost spheres. 
Notice the non-obvious dependence of $\left<\Delta F^2\right>$ on time $t$, distance $R$ and the size ratio $q$, which we next test against the results of Brownian dynamics simulations discussed next.

\section{Brownian dynamics simulations}
Brownian dynamics simulations were performed using the standard scheme developed by Ermak \cite{NOTE:platesim,ERMAK75}.  We fixed  two test colloids at a distance $R$ in a simulation box of volume $V$, and inserted $N_2$ ghost particles in the same box. The number of ghost colloids in the box is determined by the chemical potential in a hypothetical reservoir containing only ghost particles. Depending on the size ratio and the density, between 50 and 1700 small ghost particles were present in the box. The ghost particles interact with the test particles through a hard-core-like steep repulsive potential that scales as $\epsilon ((1+q)\sigma_1/r)^{50}$, where $\epsilon$ is the interaction strength and $r$ is the distance between the centers of mass of a test colloid and a host particle. The force on the particles of species 1 is rendered dimensionless by multiplying with $\sigma_2/\epsilon$.

Simulation time was divided in a sequence of intervals of width $t$. For each interval the net force $f(t)$ along the vector connecting the centers of mass of the pair of test particles was calculated. The variance of the distribution of forces we define as the fluctuation in the force. The mean depletion force is simply obtained as $\left<F\right>=\left<f( t)\right>$, while $\left<\Delta F^2\right>=\left<f(t)^2\right>-\left<f(t)\right>^2$. For small values of $t$ the distribution of forces is highly non-Gaussian, very broad and with a most probable force that is equal to zero. The distribution only becomes Gaussian for large enough $t$. We fit eq.~(\ref{eq:fluctsphere}) to the simulation results by adjusting the parameter $c$. 

\begin{figure}
\begin{center}
\includegraphics[bb=50 50 410 302]{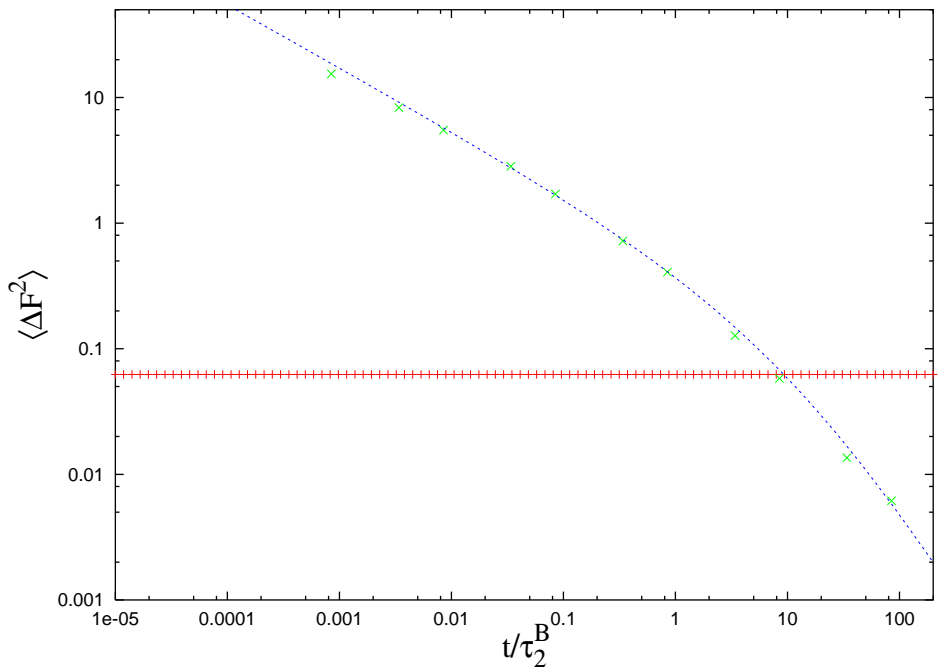}
\caption{The mean-square fluctuation of the depletion force as a function of the dimensionless time $t/\tau_2^B$ with $\tau_2^B$ the Brownian timescale of the ghost particles. The symbols indicate the simulation results and  the full curve those of  the scaling theory (eq.~(\ref{eq:fluctsphere})). The horizontal, dashed line corresponds to the average force squared $\left<F\right>^2$.}\label{fig:var_t}
\end{center}
\end{figure}

Plotted in fig.~\ref{fig:var_t} is the value of $\left<\Delta F^2\right>$ at contact, i.e., for $R=\sigma_1$, as a function of $t$ for $q=0.5$, $k_BT/\epsilon=1$ at a ghost particle density $\rho_2 \sigma_2^3= 0.0625$. The time $t$ is normalised on the Brownian time $\tau^B_2=3\pi\mu\sigma_2^3/k_BT$ of the ghost particles, which is the time required for a small particle to diffuse over its diameter. Also indicated is the fitted scaling relation eq.~(\ref{eq:fluctsphere}). For small $t$, the diffusion length is small on the scale of the linear dimension of the test particles, and therefore approaches the result for the flat plate discussed earlier. For large times the scaling crosses over to $t^{-3/2}$, approaching the time dependence of a  radially symmetric diffusion volume. Interestingly, best fits were obtained setting $c_2 =1$ although one would expect on basis of the quasi one-dimensional nature of the problem a value of $\sqrt{2}$. Note that the description is quantitative over 5 decades in time. 

Also in the case of fig.~\ref{fig:var_t} is the square-mean depletion force $\left<F\right>^2$. Clearly, $\left<\Delta F^2\right> \ll \left<F\right>^2$ only if $t\gg \tau_2^B$, i.e., for times very much larger than the diffusion time of the host particles. This means that for the case $q=0.5$, the mean depletion force cannot fully build up at the diffusion time of the guest particles. Consequently, the actual depletion force felt by these particles must be dominated by fluctuations. 

\begin{figure}
\begin{center}
\includegraphics[bb=50 50 410 302]{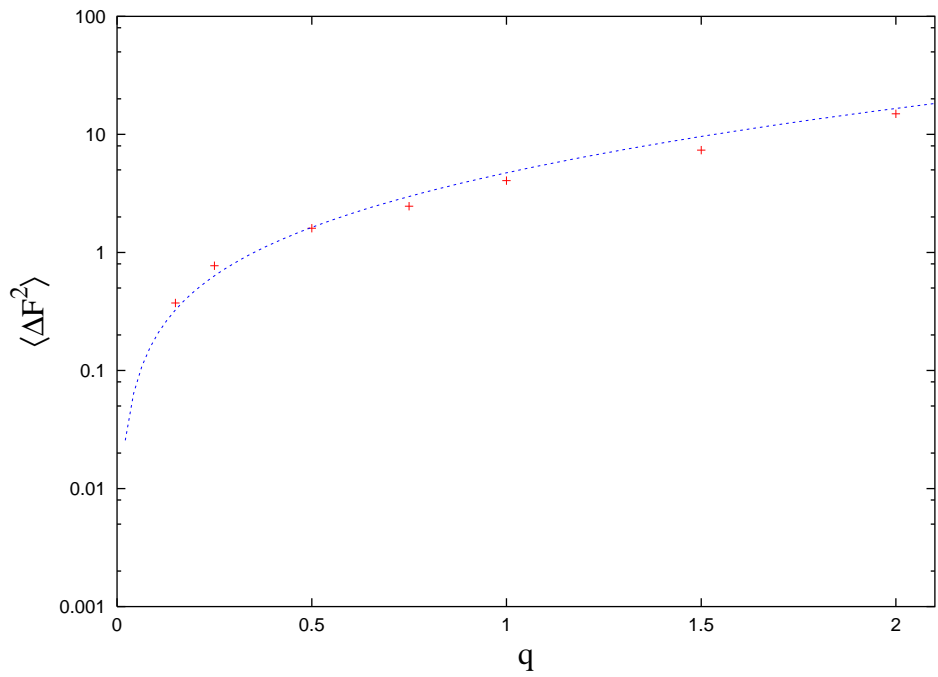}
\caption{The mean-square fluctuation in the depletion force as a function of the size ratio $q$ for $\rho_2\sigma_1^3 =0.5$, $tD_1/\sigma_1^2 =0.106$ and $R=\sigma_1$. Symbols: simulation results,  drawn curve: scaling theory (eq.~(\ref{eq:fluctsphere})).
}\label{fig:var_q}
\end{center}
\end{figure}

The dependence of the fluctuations in the depletion force on the size ratio $q$ is given in fig.~\ref{fig:var_q} for a fixed time $t D_1/\sigma_1^2=0.106$ and a fixed number density  $\rho_2\sigma_1^3=0.5$.
Shown are again the results for two test colloids at contact. Brownian dynamics simulations and scaling theory agree quantitatively and demonstrate that the fluctuations are more prominent the lager the size ratio. We were not able to obtain reliable results for $q<0.15$ as those simulations required an exceedingly small time step to maintain numerical stability.
 
\begin{figure}
\begin{center}
\includegraphics[bb=50 50 410 302]{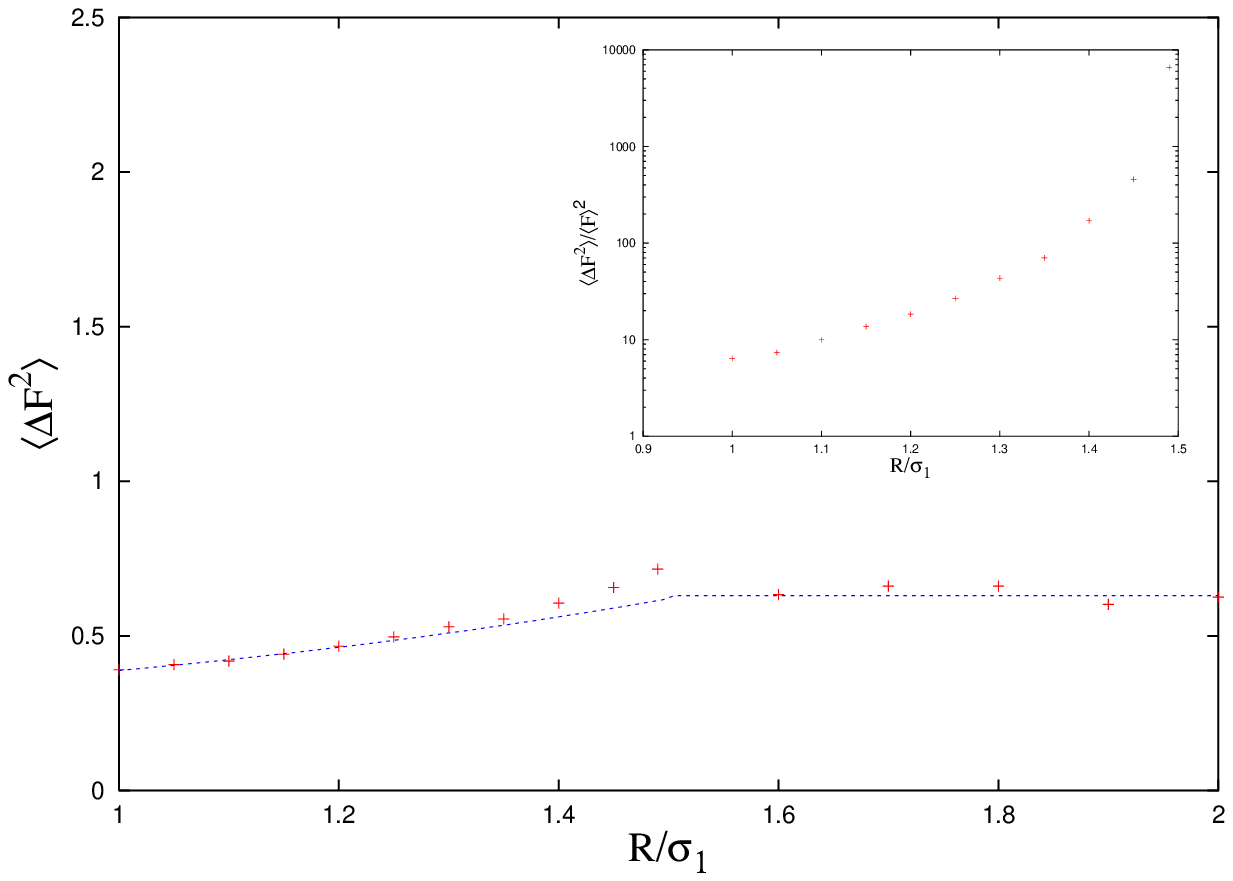}
\caption{The mean-square fluctuation in the depletion force as a function of distance $R$ between the two test particles  for $\rho_2\sigma_1^3 = 0.5$, $t D_1/\sigma_1^2=1$ and $q=0.5$. Symbols: simulation results,  drawn curve: scaling theory (eq.~(\ref{eq:fluctsphere})) fitted to the data.}\label{fig:var_R}
\end{center}
\end{figure}

Finally we show in fig.~\ref{fig:var_R} how the fluctuations vary with the distance between the particles for a given size ratio $q=0.5$, density $\rho_2\sigma_1^3=0.5$ and $t D_1/\sigma_1^2=0.106$. The absolute value of the fluctuations in the depletion force increases with the distance between the test colloids and reaches the single particle value for $R>\sigma_1+\sigma_2$. The {\it relative} fluctuations $\left<\Delta F^2\right>/\left<F\right>^2$ in fact diverge when $R\rightarrow \sigma_1+\sigma_2$ as shown in the inset to fig.~\ref{fig:var_R}. This means that fluctuations are {\it always} important in the tail of the depletion force. Note that for separations close to $R=\sigma_1+\sigma_2$, the fluctuations are slightly larger than predicted by the scaling theory. This discrepancy with scaling theory was not observed for other data sets with different $q$, $t$ and $\rho_2$, and we have no explanation for it.

\section{Conclusions}
We have shown that the temporal fluctuations in the depletion force between two guest particles in a host dispersion of colloids of different size are significant.  Our results imply that the dynamics of approach of two such colloids must be strongly influenced by these fluctuations and the slow buildup of the depletion force. It stands to reason, then, that the onset of phase separation or of gelation in mixtures of colloids must be sensitive to them. Therefore, one must be wary of treating the depletion potential as an instantaneous potential in kinetic studies \cite{BODNAR96,VERHAEGH96,MELROSE99,POON99}.

The question arises how the fluctuations discussed in this paper can be probed experimentally. One way would be to look at the dynamics of the depletion force using a tweezer experiment \cite{OHSHIMA97} or a TIRM experiment \cite{RUDHARDT98}, analysing the data in a similar fashion as we have done in our simulations. As far as we know this has not been attempted yet. Another way would be to look at the escape kinetics of a colloid pushed against a hard wall due to the presence of other colloids (similar to the study of \cite{KAPLAN94}), which should be strongly influenced by a fluctuating depletion force.

Note added in proof: Upon completion of this work we came across a preprint by Bartolo and collaborators \cite{BARTOLO02} on  fluctuations in medium-induced, Casimir-type forces between parallel plates. These Casimir-type forces are due to the coupling of the plates to a fluctuating background and are similar but not identical to the depletion forces described in this work, because the density fluctuations in our host dispersion are not scale invariant. However, they too find that the r.m.s. fluctuation in the effective interaction between them scales as $1/\sqrt{t}$ as we found in eq.~(\ref{eq:fluctplate}).

\acknowledgements
This work was funded by EU Grant No. MCFI-1999-00115 [GAV]. We thank Ard Louis, Peter Bolhuis, Bob Evans, Paul Bartlett and Gerhard Gompper for useful discussions.

\pagebreak

\begin{thebibliography}{100}

\bibitem{DIJKSTRA99}
  \Name{Dijkstra M., van Roij R.  \and Evans R.}
  \REVIEW{Phys Rev. E}{59}{1999}{5744}.

\bibitem{ASAKURA58}
  \Name{Asakura S. \and Oosawa F.}
  \REVIEW{J. Pol. Sci.}{33}{1958}{183}.

\bibitem{BODNAR96}
  \Name{ Bodn\'{a}r I., Dhont J.K.G. \and Lekkerkerker H.N.W}
  \REVIEW{J. Phys. Chem B}{100}{1996}{19614}.

\bibitem{VERHAEGH96}
 \Name{Verhaegh N.A.M., van Duijneveldt J.S., Dhont J.K.G. \and Lekkerkerker H.N.W.}
\REVIEW{Physica A}{230}{1996}{409}.

\bibitem{MELROSE99}
  \Name{Melrose J.R., Soga K.G. \and Ball R.C.}
  \REVIEW{J. Chem. Phys.}{110}{1999}{2280}.

\bibitem{POON99}
  \Name{ Poon W.C.K., Renth F., Evans R.M.L., Fairhurst D.J., Cates M.E. \and Pusey P.N.}
  \REVIEW{Phys. Rev. Lett.}{83}{1999}{1239}.

\bibitem{RUDHARDT98}
  \Name{ Rudhardt D., Bechinger C. \and Leiderer P. }
  \REVIEW{Phys. Rev. Lett.}{81}{1998}{1330}.

\bibitem{VERMA98}
  \Name{ Verma R., Crocker J.C., Lubensky T.C. \and Yodh A.G.}
  \REVIEW{Phys. Rev. Lett.}{81}{1998}{4004}.

\bibitem{LOEWEN00}
  \Name{L\"{o}wen H.}
  {Personal communication}.

\bibitem{ASAKURA54}
  \Name{Asakura S. \and Oosawa F.}
  \REVIEW{J. Chem. Phys.}{22}{1954}{1255}.

\bibitem{NOTE:platesim}
{We also tested the case of parallel plates against Brownian dynamics simulations which confirmed the scaling relation eq.~(\protect\ref{eq:fluctplate})}.

\bibitem{ERMAK75}
  \Name{Ermak D.L.}
  \REVIEW{J. Chem. Phys.}{62}{1975}{4189}.

\bibitem{OHSHIMA97}
  \Name{ Ohshima Y.N., Sakagami H., Okumoto K., Tokoyada A., Igarashi T., Shintaku K.B., Toride S., Sekino H., Kabuto K. \and Nishio I.}
  \REVIEW{Phys. Rev. Lett.}{78}{1997}{3963}.

\bibitem{KAPLAN94}
  \Name{ Kaplan P.D., Faucheux L.P. \and Libschaber A.J.}
  \REVIEW{Phys. Rev. Lett.}{73}{1994}{2793}.

\bibitem{BARTOLO02}
\Name{Bartolo D., Ajdari A., Fournier J-B., \and Golestanian R.}
{cond-mat/0208095}.

\end{thebibliography}
\end{document}